\pdfoutput=1
\documentclass[aps,prl,reprint,twocolumn,superscriptaddress,showpacs,floatfix,nobibnotes,preprintnumbers]{revtex4-1}

\usepackage{graphicx}
\usepackage{dcolumn}
\usepackage{xspace}
\usepackage[colorlinks,breaklinks]{hyperref}

\newcommand{\ppbar}{$p\bar{p}$}
\newcommand{\roots}{$\sqrt{s}$}
\newcommand{\fb}{fb$^{-1}$}
\newcommand{\gev}{GeV}

\newcommand{\gevcc}{GeV/$c^2$}

\newcommand{\et}{$E_T$}
\newcommand{\py}{\textsc{pythia}}

\newcommand{\mgg}{$m_{\gamma\gamma}$}
\newcommand{\Hgg}{\mbox{$H \rightarrow \gamma \gamma$}\xspace}

\newcommand{\Zee}{\mbox{$Z \rightarrow e^+e^-$}\xspace}
\newcommand{\Zgg}{\mbox{$Z \rightarrow \gamma \gamma$}\xspace}
\newcommand{\Zpg}{\mbox{$Z\rightarrow\pi^0\gamma$}\xspace}
\newcommand{\Zpp}{\mbox{$Z\rightarrow\pi^0\pi^0$}\xspace}
\newcommand{\BRZgg}{\mbox{$\mathcal{B}(Z \rightarrow \gamma \gamma)$}\xspace}
\newcommand{\BRZpg}{\mbox{$\mathcal{B}(Z\rightarrow\pi^0\gamma)$}\xspace}
\newcommand{\BRZpp}{\mbox{$\mathcal{B}(Z\rightarrow\pi^0\pi^0)$}\xspace}

\newcommand{\pion}{$\pi^{0}$}
\newcommand{\babar}{\mbox{{\mbox{\sl B\hspace{-0.4em} {\small\sl A}\hspace{-0.10em}\sl B\hspace{-0.4em}}} {\small\sl A\hspace{-0.02em}R}}}

\begin{document}

\preprint{FERMILAB-PUB-13-509-E}

\affiliation{Institute of Physics, Academia Sinica, Taipei, Taiwan 11529, Republic of China}
\affiliation{Argonne National Laboratory, Argonne, Illinois 60439, USA}
\affiliation{University of Athens, 157 71 Athens, Greece}
\affiliation{Institut de Fisica d'Altes Energies, ICREA, Universitat Autonoma de Barcelona, E-08193, Bellaterra (Barcelona), Spain}
\affiliation{Baylor University, Waco, Texas 76798, USA}
\affiliation{Istituto Nazionale di Fisica Nucleare Bologna, \ensuremath{^{ii}}University of Bologna, I-40127 Bologna, Italy}
\affiliation{University of California, Davis, Davis, California 95616, USA}
\affiliation{University of California, Los Angeles, Los Angeles, California 90024, USA}
\affiliation{Instituto de Fisica de Cantabria, CSIC-University of Cantabria, 39005 Santander, Spain}
\affiliation{Carnegie Mellon University, Pittsburgh, Pennsylvania 15213, USA}
\affiliation{Enrico Fermi Institute, University of Chicago, Chicago, Illinois 60637, USA}
\affiliation{Comenius University, 842 48 Bratislava, Slovakia; Institute of Experimental Physics, 040 01 Kosice, Slovakia}
\affiliation{Joint Institute for Nuclear Research, RU-141980 Dubna, Russia}
\affiliation{Duke University, Durham, North Carolina 27708, USA}
\affiliation{Fermi National Accelerator Laboratory, Batavia, Illinois 60510, USA}
\affiliation{University of Florida, Gainesville, Florida 32611, USA}
\affiliation{Laboratori Nazionali di Frascati, Istituto Nazionale di Fisica Nucleare, I-00044 Frascati, Italy}
\affiliation{University of Geneva, CH-1211 Geneva 4, Switzerland}
\affiliation{Glasgow University, Glasgow G12 8QQ, United Kingdom}
\affiliation{Harvard University, Cambridge, Massachusetts 02138, USA}
\affiliation{Division of High Energy Physics, Department of Physics, University of Helsinki, FIN-00014, Helsinki, Finland; Helsinki Institute of Physics, FIN-00014, Helsinki, Finland}
\affiliation{University of Illinois, Urbana, Illinois 61801, USA}
\affiliation{The Johns Hopkins University, Baltimore, Maryland 21218, USA}
\affiliation{Institut f\"{u}r Experimentelle Kernphysik, Karlsruhe Institute of Technology, D-76131 Karlsruhe, Germany}
\affiliation{Center for High Energy Physics: Kyungpook National University, Daegu 702-701, Korea; Seoul National University, Seoul 151-742, Korea; Sungkyunkwan University, Suwon 440-746, Korea; Korea Institute of Science and Technology Information, Daejeon 305-806, Korea; Chonnam National University, Gwangju 500-757, Korea; Chonbuk National University, Jeonju 561-756, Korea; Ewha Womans University, Seoul, 120-750, Korea}
\affiliation{Ernest Orlando Lawrence Berkeley National Laboratory, Berkeley, California 94720, USA}
\affiliation{University of Liverpool, Liverpool L69 7ZE, United Kingdom}
\affiliation{University College London, London WC1E 6BT, United Kingdom}
\affiliation{Centro de Investigaciones Energeticas Medioambientales y Tecnologicas, E-28040 Madrid, Spain}
\affiliation{Massachusetts Institute of Technology, Cambridge, Massachusetts 02139, USA}
\affiliation{University of Michigan, Ann Arbor, Michigan 48109, USA}
\affiliation{Michigan State University, East Lansing, Michigan 48824, USA}
\affiliation{Institution for Theoretical and Experimental Physics, ITEP, Moscow 117259, Russia}
\affiliation{University of New Mexico, Albuquerque, New Mexico 87131, USA}
\affiliation{The Ohio State University, Columbus, Ohio 43210, USA}
\affiliation{Okayama University, Okayama 700-8530, Japan}
\affiliation{Osaka City University, Osaka 558-8585, Japan}
\affiliation{University of Oxford, Oxford OX1 3RH, United Kingdom}
\affiliation{Istituto Nazionale di Fisica Nucleare, Sezione di Padova, \ensuremath{^{jj}}University of Padova, I-35131 Padova, Italy}
\affiliation{University of Pennsylvania, Philadelphia, Pennsylvania 19104, USA}
\affiliation{Istituto Nazionale di Fisica Nucleare Pisa, \ensuremath{^{kk}}University of Pisa, \ensuremath{^{ll}}University of Siena, \ensuremath{^{mm}}Scuola Normale Superiore, I-56127 Pisa, Italy, \ensuremath{^{nn}}INFN Pavia, I-27100 Pavia, Italy, \ensuremath{^{oo}}University of Pavia, I-27100 Pavia, Italy}
\affiliation{University of Pittsburgh, Pittsburgh, Pennsylvania 15260, USA}
\affiliation{Purdue University, West Lafayette, Indiana 47907, USA}
\affiliation{University of Rochester, Rochester, New York 14627, USA}
\affiliation{The Rockefeller University, New York, New York 10065, USA}
\affiliation{Istituto Nazionale di Fisica Nucleare, Sezione di Roma 1, \ensuremath{^{pp}}Sapienza Universit\`{a} di Roma, I-00185 Roma, Italy}
\affiliation{Mitchell Institute for Fundamental Physics and Astronomy, Texas A\&M University, College Station, Texas 77843, USA}
\affiliation{Istituto Nazionale di Fisica Nucleare Trieste, \ensuremath{^{qq}}Gruppo Collegato di Udine, \ensuremath{^{rr}}University of Udine, I-33100 Udine, Italy, \ensuremath{^{ss}}University of Trieste, I-34127 Trieste, Italy}
\affiliation{University of Tsukuba, Tsukuba, Ibaraki 305, Japan}
\affiliation{Tufts University, Medford, Massachusetts 02155, USA}
\affiliation{University of Virginia, Charlottesville, Virginia 22906, USA}
\affiliation{Waseda University, Tokyo 169, Japan}
\affiliation{Wayne State University, Detroit, Michigan 48201, USA}
\affiliation{University of Wisconsin, Madison, Wisconsin 53706, USA}
\affiliation{Yale University, New Haven, Connecticut 06520, USA}

\author{T.~Aaltonen}
\affiliation{Division of High Energy Physics, Department of Physics, University of Helsinki, FIN-00014, Helsinki, Finland; Helsinki Institute of Physics, FIN-00014, Helsinki, Finland}
\author{S.~Amerio\ensuremath{^{jj}}}
\affiliation{Istituto Nazionale di Fisica Nucleare, Sezione di Padova, \ensuremath{^{jj}}University of Padova, I-35131 Padova, Italy}
\author{D.~Amidei}
\affiliation{University of Michigan, Ann Arbor, Michigan 48109, USA}
\author{A.~Anastassov\ensuremath{^{v}}}
\affiliation{Fermi National Accelerator Laboratory, Batavia, Illinois 60510, USA}
\author{A.~Annovi}
\affiliation{Laboratori Nazionali di Frascati, Istituto Nazionale di Fisica Nucleare, I-00044 Frascati, Italy}
\author{J.~Antos}
\affiliation{Comenius University, 842 48 Bratislava, Slovakia; Institute of Experimental Physics, 040 01 Kosice, Slovakia}
\author{G.~Apollinari}
\affiliation{Fermi National Accelerator Laboratory, Batavia, Illinois 60510, USA}
\author{J.A.~Appel}
\affiliation{Fermi National Accelerator Laboratory, Batavia, Illinois 60510, USA}
\author{T.~Arisawa}
\affiliation{Waseda University, Tokyo 169, Japan}
\author{A.~Artikov}
\affiliation{Joint Institute for Nuclear Research, RU-141980 Dubna, Russia}
% ...

\collaboration{CDF Collaboration}
\altaffiliation[With visitors from]{
\ensuremath{^{a}}University of British Columbia, Vancouver, BC V6T 1Z1, Canada,
\ensuremath{^{b}}Istituto Nazionale di Fisica Nucleare, Sezione di Cagliari, 09042 Monserrato (Cagliari), Italy,
\ensuremath{^{c}}University of California Irvine, Irvine, CA 92697, USA,
\ensuremath{^{d}}Institute of Physics, Academy of Sciences of the Czech Republic, 182~21, Czech Republic,
\ensuremath{^{e}}CERN, CH-1211 Geneva, Switzerland,
\ensuremath{^{f}}Cornell University, Ithaca, NY 14853, USA,
\ensuremath{^{g}}University of Cyprus, Nicosia CY-1678, Cyprus,
\ensuremath{^{h}}Office of Science, U.S. Department of Energy, Washington, DC 20585, USA,
\ensuremath{^{i}}University College Dublin, Dublin 4, Ireland,
\ensuremath{^{j}}ETH, 8092 Z\"{u}rich, Switzerland,
\ensuremath{^{k}}University of Fukui, Fukui City, Fukui Prefecture, Japan 910-0017,
\ensuremath{^{l}}Universidad Iberoamericana, Lomas de Santa Fe, M\'{e}xico, C.P. 01219, Distrito Federal,
\ensuremath{^{m}}University of Iowa, Iowa City, IA 52242, USA,
\ensuremath{^{n}}Kinki University, Higashi-Osaka City, Japan 577-8502,
\ensuremath{^{o}}Kansas State University, Manhattan, KS 66506, USA,
\ensuremath{^{p}}Brookhaven National Laboratory, Upton, NY 11973, USA,
\ensuremath{^{q}}Queen Mary, University of London, London, E1 4NS, United Kingdom,
\ensuremath{^{r}}University of Melbourne, Victoria 3010, Australia,
\ensuremath{^{s}}Muons, Inc., Batavia, IL 60510, USA,
\ensuremath{^{t}}Nagasaki Institute of Applied Science, Nagasaki 851-0193, Japan,
\ensuremath{^{u}}National Research Nuclear University, Moscow 115409, Russia,
\ensuremath{^{v}}Northwestern University, Evanston, IL 60208, USA,
\ensuremath{^{w}}University of Notre Dame, Notre Dame, IN 46556, USA,
\ensuremath{^{x}}Universidad de Oviedo, E-33007 Oviedo, Spain,
\ensuremath{^{y}}CNRS-IN2P3, Paris, F-75205 France,
\ensuremath{^{z}}Universidad Tecnica Federico Santa Maria, 110v Valparaiso, Chile,
\ensuremath{^{aa}}The University of Jordan, Amman 11942, Jordan,
\ensuremath{^{bb}}Universite catholique de Louvain, 1348 Louvain-La-Neuve, Belgium,
\ensuremath{^{cc}}University of Z\"{u}rich, 8006 Z\"{u}rich, Switzerland,
\ensuremath{^{dd}}Massachusetts General Hospital, Boston, MA 02114 USA,
\ensuremath{^{ee}}Harvard Medical School, Boston, MA 02114 USA,
\ensuremath{^{ff}}Hampton University, Hampton, VA 23668, USA,
\ensuremath{^{gg}}Los Alamos National Laboratory, Los Alamos, NM 87544, USA,
\ensuremath{^{hh}}Universit\`{a} degli Studi di Napoli Federico I, I-80138 Napoli, Italy
}
\noaffiliation

\title{First Search for Exotic \boldmath $Z$ Boson Decays into Photons and Neutral Pions \\ in Hadron Collisions}

\date{March 20, 2014}

\begin{abstract}
A search for forbidden and exotic $Z$ boson decays in the diphoton 
mass spectrum is presented for the first time in hadron collisions, 
based on data corresponding to 10.0~\fb\ of integrated luminosity from 
proton-antiproton collisions at \roots~=~1.96~TeV collected by the CDF 
experiment. No evidence of signal is observed, and 95\% credibility 
level Bayesian upper limits are set on the branching ratios of decays 
of the $Z$ boson to a photon and neutral pion (which is detected as a 
photon), a pair of photons, and a pair of neutral pions. The observed 
branching ratio limits are 2.01~$\times$~10$^{-5}$ for \Zpg, 
1.46~$\times$~10$^{-5}$ for \Zgg, and 1.52~$\times$~10$^{-5}$ for 
\Zpp. The \Zpg and \Zgg limits improve the most stringent results from 
other experiments by factors of 2.6 and 3.6, respectively. The \Zpp 
branching ratio limit is the first experimental result on this decay. 
\end{abstract}


\maketitle

\vspace*{2ex}

Properties of the $W$ and $Z$ bosons have been studied extensively in 
collider experiments. Most knowledge about these particles from hadron 
colliders, however, comes from decays involving leptons~\cite{PDG}. 
Although the hadronic decay modes of the $W$ and $Z$ bosons dominate, 
identifying a $W$ or $Z$ boson resonance from the quark-antiquark 
final state is challenging at hadron colliders due to large 
backgrounds and poor jet energy resolution. It is therefore appealing 
to consider rare $V\rightarrow P + \gamma$ decays, where $V$ is a weak 
vector boson and $P$ is a pseudoscalar meson. To complement the search 
for $W^{\pm} \rightarrow \pi^{\pm}\gamma$ decays performed previously 
at CDF~\cite{WpgCDF}, we present a new search for \Zpg\ decays. This 
decay offers a unique opportunity to reconstruct the $Z$ boson from an 
isolated hadron~\cite{ZpigamDecay}, rather than from its decay to either 
a lepton pair or a quark-antiquark pair, and it is sensitive to couplings of 
the $Z$ boson to quarks and the photon~\cite{WZPiGamDecays}. 

The standard model (SM) 
branching ratio prediction for \Zpg ranges between $10^{-12}$ and 
$10^{-9}$~\cite{Deshpande,*Hikasa,*Manohar,*Pham,*West,*Young,*Schroder,*Chatterjee,*Maitra,*Micu}, 
which is too small for such a process to be detected at the Tevatron. Nonetheless, 
evidence of a signal may indicate physics beyond the SM, and the 
absence of a signal would improve the current experimental upper 
bounds on the \Zpg branching ratio. Furthermore, an experimental limit 
on \BRZpg would improve upper bounds for the pion transition form 
factor, which describes the $\pi^{0}\to\gamma\gamma^*$ transition. 
Measurements of this factor by the \babar\ experiment were found to be 
higher than the predictions of perturbative quantum chromodynamics for 
the momentum transfer ($Q^2$) range 15--34~GeV$^2$~\cite{BABAR}. 
Theoretical efforts since then have attempted to explain this 
discrepancy, some using experimental \BRZpg upper limits as an 
additional constraint at higher $Q^2$~\cite{Arriola,*Pham2011}. 

In this search for \Zpg\ decays, the $\pi^{0}$ and the $\gamma$ have 
similar experimental signatures in the CDF II detector.  It is therefore 
natural to extend the search to include the decays 
\Zgg and \Zpp, which are quantum mechanically forbidden due to the 
conservation of angular momentum applied to identical final-state 
particles (thus violating the spin-statistics theorem)~\cite{ZpigamDecay,Landau,*Yang}.  
The \Zgg decay was studied in the past by experiments at the Large 
Electron-Positron (LEP) collider~\cite{L3,Delphi,Opal}, and the resulting 
limits on the \Zgg branching ratio have been used~\cite{ZggBose,*Gninenko} 
to constrain possible small violations of Bose-Einstein statistics, complementing 
other similar analyses~\cite{PhysRevD.39.2032,*Demille,*Hilborn}. 
The \Zgg decay is also considered a promising process for discovery of 
physics beyond the SM that allows for noncommutative 
space-time~\cite{Hinchliffe,*Behr,*Duplancic,*Buric,*Trampetic,*Horvat:2012vn}. 

The most stringent existing experimental limits on the \Zpg and \Zgg 
branching ratios come from LEP experiments \cite{PDG}. Specifically, 
the L3 experiment set a 95\% confidence level limit of $5.2 \times 10^{-5}$ 
on the branching ratio of both the \Zpg and \Zgg decay 
modes~\cite{L3}. Though the experimental search for these decays is 
challenging due to their small branching ratios, the abundance of $Z$ 
bosons produced in high-energy hadron collisions at 
CDF~\cite{Aaltonen2010232} (about 4 times the number of $Z$ bosons 
produced at LEP \cite{zreslep}) allows an improvement of the existing 
limits. The intensive search for the SM Higgs boson in the diphoton 
decay mode~\cite{cdfHgg2012,*cdfHgg2012b,*ThesisBland} 
has led to dedicated data analysis techniques that can be 
directly applied to \Zpg, \Zgg, and \Zpp searches. No limits on the 
branching ratios of these decay modes from experiments in hadron 
collisions have been reported to date. In this Letter, we present a 
search for \Zpg, \Zgg, and \Zpp decays using the full CDF diphoton 
data set, corresponding to an integrated luminosity of 10.0~fb$^{-1}$. 

The CDF II detector~\cite{Acosta:2004yw} is used to identify photon 
candidate events produced in \ppbar\ collisions at \roots~= 1.96~TeV. 
The silicon vertex tracker~\cite{Aaltonen2013153} and the central 
outer tracker~\cite{Affolder:2003ep}, contained within a 1.4~T axial 
magnetic field, measure the trajectories (tracks) of charged particles 
and determine their momenta. Particles that pass through the tracking 
volume reach the electromagnetic (EM) and hadronic 
calorimeters~\cite{ccal, *wha, *Albrow:2001jw}, which are divided into 
two regions: central ($|\eta|<1.1$)~\cite{CDFcoord} and forward 
($1.1<|\eta|<3.6$). The EM calorimeters contain fine-grained shower 
maximum detectors~\cite{Apollinari:1998bg}, which measure the shower 
shape and centroid position in the two dimensions transverse to the 
direction of the shower development. 

Events having two isolated EM showers with no associated tracks are 
selected by a three-level on-line event-selection system (trigger) that 
requires an isolated cluster of energy deposited in the EM calorimeter 
with transverse energy \et~$>25$~\gev~\cite{EtDef}. The trigger 
efficiency for events accepted into the final sample is determined 
from simulated events and found to be 
(99.8 $\pm$ 1.0)\%~\cite{cdfHgg2012,*cdfHgg2012b,*ThesisBland}. 

A neural network (NN) technique is used to reconstruct the two 
highest-$E_T$ photon candidates in the event with $E_T>15$~GeV/$c$ in 
the fiducial region of the central EM calorimeter ($|\eta|< 1.05$). 
This selection was developed in the SM \Hgg analysis at 
CDF~\cite{cdfHgg2012,*cdfHgg2012b,*ThesisBland}. The NN selection is 
optimized for the identification of high-$E_T$ central photons and the 
rejection of the dominant background from jets. This selection also 
identifies neutral pions from the $Z$ boson signal with high 
efficiency. For \Zpg or \Zpp decays, the $\pi^{0}$ is isolated (not 
contained in a jet) and decays about 99\% of the time into a pair of 
photons. Because of the high momentum (on average 45 GeV/$c$) of the 
$\pi^{0}$ from a $Z$ boson decay, the photon pair is usually produced 
with a sufficiently narrow opening angle such that the two photons 
appear in the central shower-maximum detector and the central EM 
calorimeter as a single EM shower. Neutral pions from a $Z$ decay then 
have nearly the same signature as an isolated photon, with only a 
slightly smaller photon identification efficiency. Therefore, any 
evidence of a signal in the diphoton data could indicate the 
identification of an isolated photon and a neutral pion, a pair of 
isolated photons, or a pair of neutral pions. 

Simulated events from the {\sc pythia} Monte Carlo (MC) 
generator~\cite{PYTHIA} are used to predict the diphoton mass (\mgg) 
shape and the product of the efficiency and detector acceptance 
($\epsilon$A) of the signal. {\sc pythia} version 6.2.16 is used with 
the CTEQ5L~\cite{CTEQ5L} parton distribution function set and an 
underlying event configuration tuned to CDF 
data~\cite{TUNEA,*Aaltonen:2010rm}. We generate $Z$ bosons inclusively 
with {\sc pythia} using a relativistic Breit-Wigner mass distribution 
for the production and an angular distribution for the decay in the 
$Z$ rest frame corresponding to the spin transition 
$1\rightarrow(1/2,1/2)$. Since {\sc pythia} does not model the \Zpg, 
\Zgg, and \Zpp decay modes, the decay products in {\sc pythia} are 
taken to be neutrinos (assumed massless) and afterward treated 
artificially as photons in the detector simulation; the events are 
then weighted by factors that transform the original angular 
distribution to one appropriate for the spin transitions 
$1\rightarrow(0,1)$ and $1\rightarrow(1,1)$, which correspond to the 
\Zpg and \Zgg decay modes, respectively. The weighting factors are 1 
for \Zpg decays and $(a-\cos^{2}\vartheta)/(b+\cos^{2}\vartheta)$ for 
\Zgg decays, where $\vartheta$ is the angle between the proton 
momentum and the momentum of each decay product in the $Z$ boson rest 
frame. For \Zpp decays, we use the same weighting factor as for \Zgg 
decays, assuming a nonzero orbital angular momentum for the pion pair 
in the final state. The constants $a=1.16$ and $b=1.32$ are derived 
from the {\sc pythia} spin-density matrix of the $Z$ boson, which 
accurately describes the angular distributions of \Zee decays measured 
with the CDF~II detector~\cite{Sakumoto:2011a4}. The events are then 
processed through the CDF~II detector and trigger simulation and event 
reconstruction software~\cite{cdfSim}. The final weighted samples 
simulate the $Z$ boson production and decay properties, including 
polarization, and properly model properties that affect the detector 
acceptance. 

\begin{figure}[tb]
\begin{center}
\includegraphics[width=1.0\linewidth]{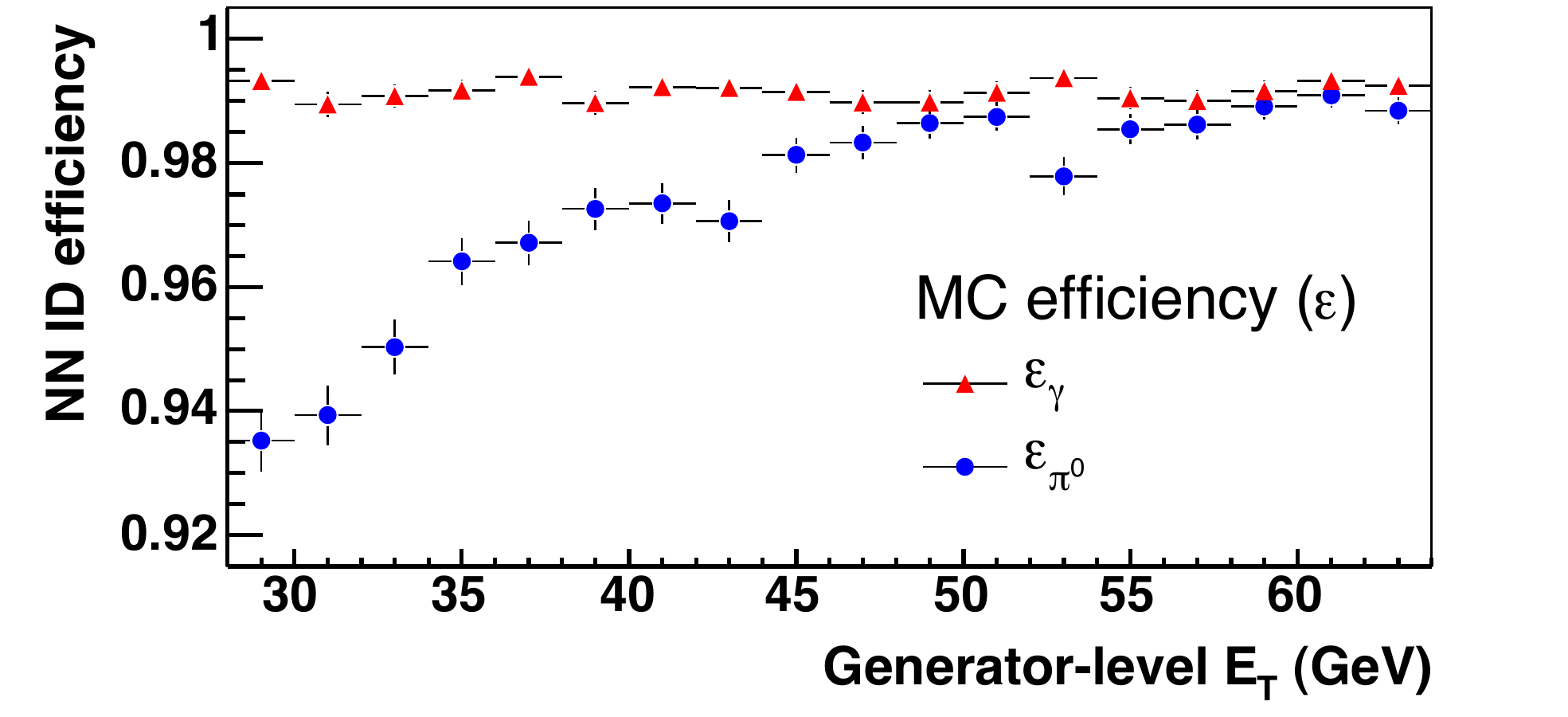}
\end{center}
\caption{Comparison of the $\gamma$ and $\pi^{0}$ NN selection 
efficiencies, obtained using the special-purpose MC event generator. 
The efficiency is shown as a function of the generator-level $E_T$. 
\label{fig:PhoIDEff}}
\end{figure}

To correctly model the efficiency for reconstructing \Zpg and \Zpp 
decays, an additional weight is applied to the simulated signal events 
based on the ratio of the $\pi^0$ to $\gamma$ NN selection 
efficiencies. Because of the small opening angle of the photon pair from the 
decay of a $\pi^{0}$, the NN selection efficiency for an isolated $\pi^{0}$ 
is slightly smaller than that for an isolated $\gamma$. These NN 
efficiencies, shown in Fig.~\ref{fig:PhoIDEff}, are obtained using a 
special-purpose MC event generator that produces single particles with 
a flat $E_T$ spectrum and does not account for the underlying event. 
The $\pi^{0}/\gamma$ efficiency weights are applied to the simulated 
events on an event-by-event basis as a function of the generated 
$E_T$, once for the \Zpg decay and twice for the \Zpp decay. As the 
decay products from the $Z$ boson have a generated $E_T$ distribution 
that peaks around 45~GeV, the weights reduce the overall 
reconstruction efficiency for \Zpg (\Zpp) decays by about 2\% (4\%) 
relative to \Zgg decays. 

In addition to the angle and efficiency weights applied to the 
simulated signal samples, events from these simulated samples are 
further weighted such that the distribution of the number of 
reconstructed vertices for the sample is the same as that observed in 
the diphoton data. This has a negligible effect on the diphoton mass 
resolution. We then assume the same \mgg\ shape for each signal 
process, which is justified because the mean and width of the 
invariant mass for each process agree to within 1\%. The resulting $Z$ 
boson signal has a natural width of 2.5~GeV, and is smeared by an 
additional 2.4~GeV due to the EM calorimeter resolution. A signal 
region of $80<m_{\gamma\gamma}<102$~\gevcc\ is defined, which contains 
90\% of the simulated signal events. Corrections to the NN selection 
efficiency due to imperfections in the detector simulation are also 
applied. These corrections are derived using electrons from \Zee 
decays by comparing the selection efficiencies obtained from the 
detector simulation to the selection efficiencies measured in the 
data~\cite{Aaltonen:2009ga}. The products of the efficiency and detector 
acceptance ($\epsilon A$) for the diphoton selection in the 80--102~\gevcc\ 
signal region, along with the associated systematic uncertainties,
are (5.67~$\pm$ 0.42)\%, (7.80~$\pm$ 0.56)\%, and (7.50~$\pm$ 0.61)\% for 
the \Zpg, \Zgg, and \Zpp decay modes, respectively. The \Zpg decay has a 
lower value due primarily to its different angular distribution compared to 
that of \Zpp and \Zpg decays.

\begin{figure*}[tb]
\begin{center}
\includegraphics[width=0.49\linewidth]{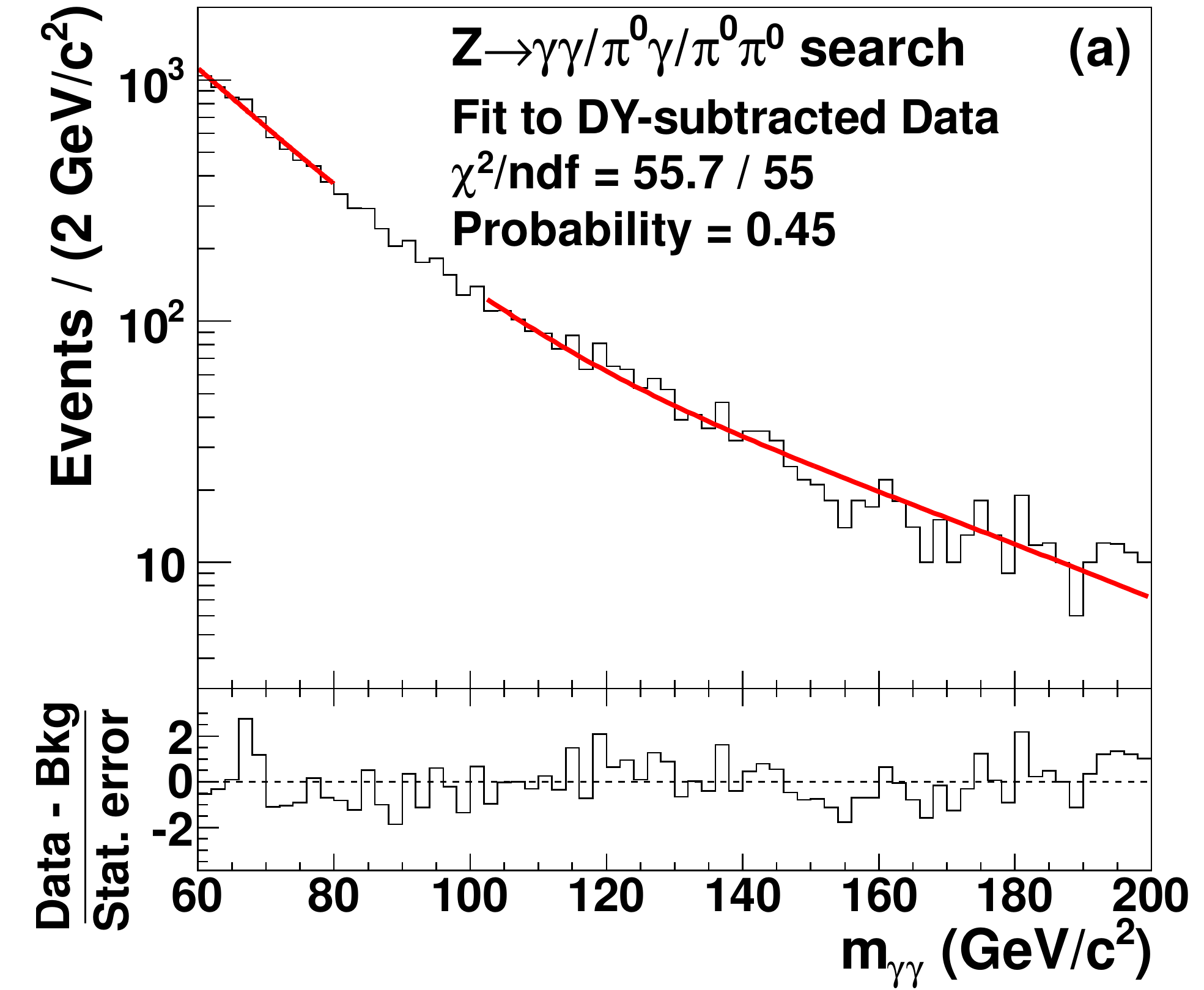}
\includegraphics[width=0.49\linewidth]{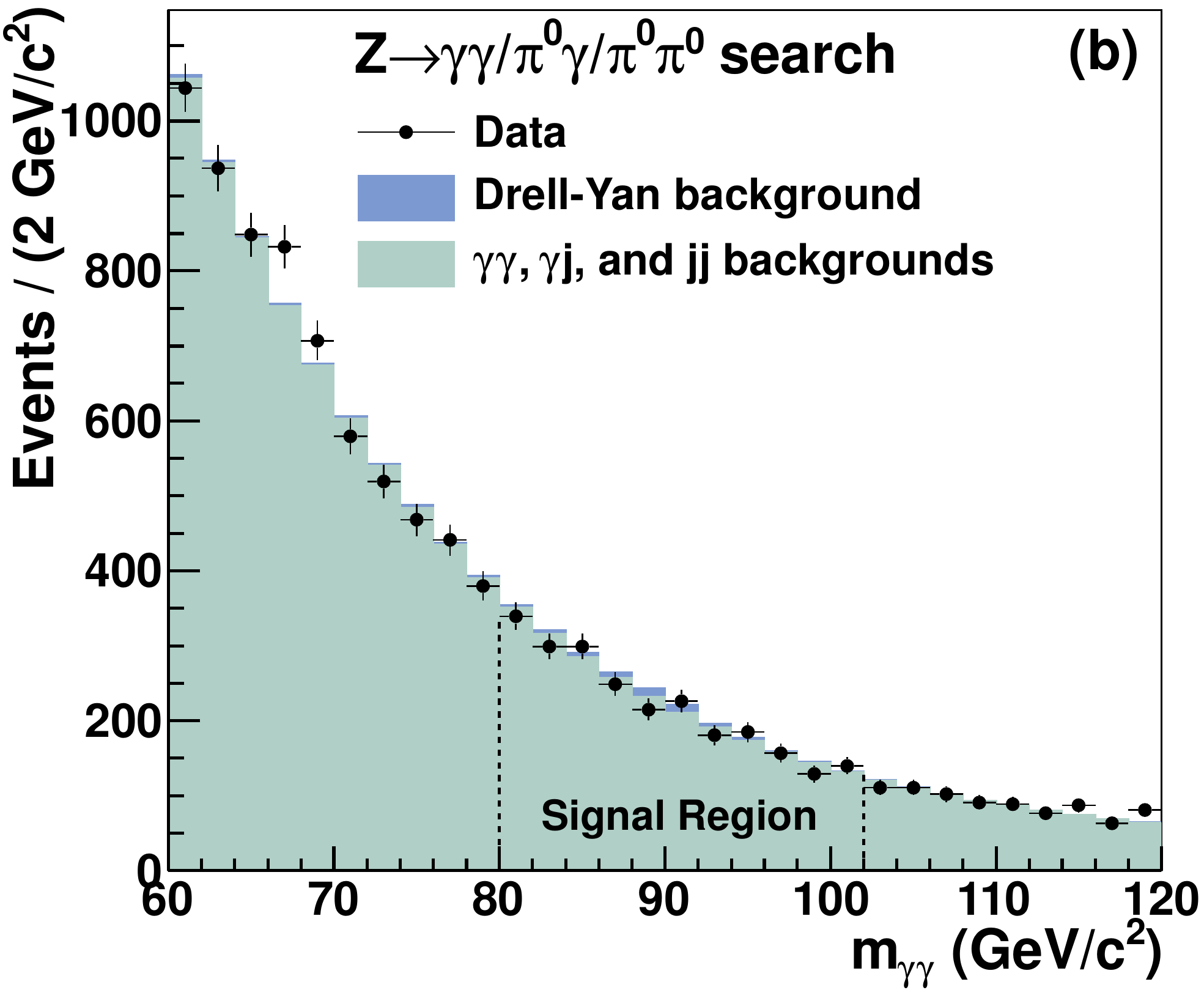}\\[3ex]
\includegraphics[width=0.49\linewidth]{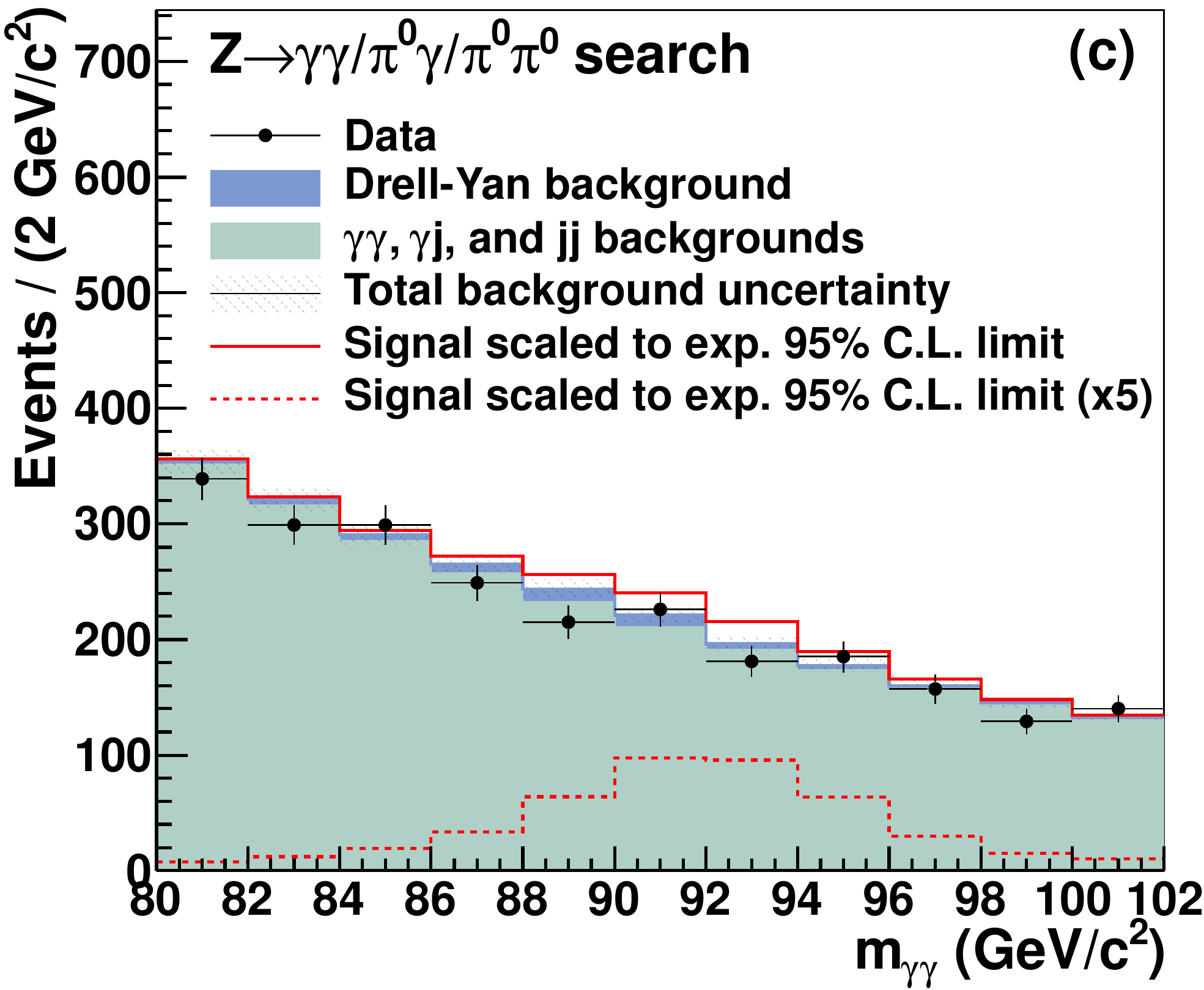}
\end{center}
\caption{(a) The diphoton mass distribution of the DY-subtracted data with 
the fit results overlaid. The gap in the fit function represents the 
excluded signal region. The lower subfigure shows the difference 
between the data and background predictions for the full 
60--200~\gevcc\ region, divided by the statistical uncertainty from 
the background prediction. (b) The diphoton mass distribution of the 
Drell-Yan and nonresonant ($\gamma\gamma$, $\gamma j$, and $jj$) 
backgrounds, compared to the data in the region 60--120~\gevcc, which 
includes part of the fit region along with the 80--102~\gevcc\ signal region. 
(c) The same distributions as (b), but for the signal region alone. Overlaid 
on the background is the predicted signal shape (solid line), 
scaled to the expected 95\% C.L. upper limit. For better visibility, 
the same signal shape is also shown along the horizontal axis multiplied 
by 5 (dashed line). 
\label{fig:Mgg}}
\end{figure*}

The decay of the $Z$ boson into \pion$\gamma$, $\gamma\gamma$, or 
\pion\pion\ would appear as a narrow peak in the invariant mass 
distribution of the two reconstructed photons. The dominant 
backgrounds are a resonant Drell-Yan (DY) component (about 2\%), which 
consists primarily of $Z \to e^+e^-$ decays~\cite{dyd}, and a nonresonant 
component (about 98\%). An inclusive \Zee\ {\sc pythia} MC sample is 
used to model the DY background component, in which a pair of 
electrons pass the diphoton selection with $\epsilon A = (1.11 \pm 
0.02)~\times~10^{-5}$ in the signal region. The expected DY yield in 
this region is $55\pm5$ events. The dominant, nonresonant background 
is composed mostly (about 2/3) of events where one or two jets are 
identified as a reconstructed photon, denoted respectively as the 
$\gamma j$ and $jj$ backgrounds. The remainder of the nonresonant 
background arises from QCD processes that produce two real photons in 
the hard interaction, denoted as the $\gamma\gamma$ background. The 
total nonresonant background prediction is estimated from a fit to 
the DY-subtracted data using a binned log-likelihood method. The data 
are fit to an exponential multiplied by a second-degree polynomial in 
the region 60--200~\gevcc, with the signal region window excluded from 
the fit. The predicted background in the signal region is obtained by 
interpolating the fit into this region. The fit to the data is shown 
in Fig.~\hyperref[fig:Mgg]{\ref{fig:Mgg}(a)}, where $2452\pm66$ events are 
expected in the signal region. The uncertainty on this background (2.7\%) arises 
from the propagation of the parameter uncertainties in the fit to the event 
yield in the signal region. Figure~\hyperref[fig:Mgg]{\ref{fig:Mgg}(b)} compares 
the Drell-Yan and nonresonant backgrounds to the data for the region 
$60<m_{\gamma\gamma}<120$~\gevcc. An excess of events is visible near 
\mgg~$= 67$~\gevcc. Although this mass region is not the subject of 
this search, we verify that the statistical significance of this 
excess is less than 2$\sigma$ after accounting for the look-elsewhere 
effect~\cite{LEE}. Table~\ref{Tab:EventYields} provides the background 
and data yields in the signal region. 

\begin{table}[h]
\caption{Number of events expected in the 80--102~\gevcc\ signal 
region for the Drell-Yan and nonresonant ($\gamma\gamma$, $\gamma j$, 
and $jj$) backgrounds, along with systematic uncertainties. The total 
background and the corresponding number of observed events in the data 
are also given.} \label{Tab:EventYields}
\begin{center}
\begin{tabular}{l@{\hskip 7mm}c} 
\hline
\hline
Background process  &  \multicolumn{1}{c}{Number of events} \\
\hline 
Drell-Yan & \, $55 \pm 5 $\\
$\gamma\gamma$, $\gamma j$, and $jj$ & $2452 \pm 66$ \\
\hline
Total background & $2507 \pm 66$\\
Data & \hspace{-7.5mm}2419\\
\hline
\hline
\end{tabular}
\end{center}
\end{table}

The dominant uncertainty in the analysis is from the uncertainty in 
the nonresonant background prediction, as described above. The 
largest systematic uncertainties on the expected number of signal 
events arise from the integrated luminosity measurement (6\%) and the 
measured $Z$ boson cross section (6\%)~\cite{Aaltonen2010232}, which 
are also uncertainties applied to the \Zee MC modeling. Uncertainties 
on the efficiency and detector acceptance of the signal come from 
varying the parton distribution functions used in \py\ 
(5\%)~\cite{CTEQ6.1M,*Weights} and the parameters controlling the 
amount of initial- and final-state radiation from the parton shower 
model of \py\ ($3\%$)~\cite{IFSRUnc}; they also include uncertainties 
in the NN photon selection efficiency (4\%), the $\pi^{0}/\gamma$ 
efficiency weight (2\% per $\pi^{0}$), the trigger efficiency (1\%), 
and the EM energy scale (less than 1\%). Uncertainties on the 
efficiency and detector acceptance of \Zee events come from the 
electron misidentification rate (2\%) and the trigger efficiency 
(1\%); other uncertainties on $\epsilon A$ for \Zee events are 
negligible. 

No evidence of a narrow peak or any other anomalous structure is 
visible in the signal region of the diphoton mass spectrum, and we 
calculate upper limits on each of the \Zpg, \Zgg, and \Zpp branching 
ratios. Because the three decay modes are nearly indistinguishable in 
the detector, we set a limit on the branching ratio of each decay mode 
independently, assuming the other decay modes are not present. We 
calculate a Bayesian credibility level (C.L.) limit based on a Poisson binned 
likelihood constructed from each bin in the signal region 
(80--102~GeV/$c^{2}$, with a 2~GeV/$c^{2}$ bin width) of the 
background, data, and signal mass distributions. The background 
distribution is scaled to the expected number of events in the signal 
region and the signal distribution is scaled based on the experimental 
total cross section for $Z$ boson production 
(7353~pb~\cite{Aaltonen2010232}), the integrated luminosity of the 
data sample, and the value of $\epsilon A$ for the signal region. We 
assume that the signal branching ratio can have any non-negative value 
with equal prior probability. We integrate over the systematic 
uncertainties, each assumed to be described by a Gaussian prior 
probability density truncated to avoid unphysical values. A 95\% C.L. 
limit is determined such that 95\% of the posterior density for the 
branching ratio falls below the limit~\cite{PDG}. For comparison, 
thousands of simulated experiments are also performed, based on 
expected backgrounds. The median limit for these trials is the 
expected limit, and the region where 68\% (95\%) of the trials lie 
around the median is the $1\sigma$ (2$\sigma$) expected region. 
Table~\ref{Tab:limits} provides the expected and observed 95\% C.L. 
limits on the \Zpg, \Zgg, and \Zpp branching ratios. 
Figure~\hyperref[fig:Mgg]{\ref{fig:Mgg}(c)} shows the \mgg\ distributions 
for the data and background in the signal region, with the signal shape 
scaled to the expected limit. 

\begin{table}[h]
\caption{Upper limits at 95\% C.L. on the \Zgg, \Zpg, and \Zpp 
branching ratios. Each upper limit is based on the assumption that the 
$Z$ boson decays through the corresponding process.}
\label{Tab:limits}
\begin{center}
\begin{tabular}{l@{\hskip 4.5mm}ccccc@{\hskip 3.5mm}c} 
\hline
\hline
& & \multicolumn{5}{c}{95\% C.L. limits ($\times10^{-5}$)}   \\ \cline{2-7}
\multicolumn{1}{c@{\hskip 4.5mm}}{Signal} & \multicolumn{5}{c}{Expected}  &      \\
\multicolumn{1}{c@{\hskip 4.5mm}}{process} & $-2\sigma$ & $-1\sigma$ & Median & $+1\sigma$ & $+2\sigma$ & Observed \\
\hline
\BRZpg & \, $1.24$ & \, $1.68$ & $2.34$ & $3.29$ & \, $4.54$ & $2.01$   \\
\BRZgg & \, $0.87$ & \, $1.21$ & $1.72$ & $2.40$ & \, $3.34$ & $1.46$    \\
\BRZpp & \, $0.94$ & \, $1.28$ & $1.76$ & $2.52$ & \, $3.39$ & $1.52$    \\
\hline
\hline
\end{tabular}
\end{center}
\end{table}

This Letter presents the most sensitive search to date for forbidden 
and exotic decays of the $Z$ boson to a neutral pion and photon, a 
pair of photons, and a pair of neutral pions. We search for a narrow 
resonance in the diphoton mass spectrum around 91~\gevcc\ using the 
full diphoton data collected by the CDF II detector at the Tevatron. 
No significant evidence of a resonance is found in the data. Upper 
bounds on the signal branching ratios are determined, resulting in 
observed 95\% C.L. limits of 2.01~$\times$~10$^{-5}$, 
1.46~$\times$~10$^{-5}$, and 1.52~$\times$~10$^{-5}$ for \Zpg, \Zgg, 
and \Zpp, respectively. The \Zpg and \Zgg limits are more sensitive by 
factors of 2.6 and 3.6, respectively, than the most stringent limits 
available. The \Zpp limit is the first experimental result on this 
decay.

\begin{acknowledgments}
We thank the Fermilab staff and the technical staffs of the 
participating institutions for their vital contributions. This work 
was supported by the U.S. Department of Energy and National Science 
Foundation; the Italian Istituto Nazionale di Fisica Nucleare; the 
Ministry of Education, Culture, Sports, Science and Technology of 
Japan; the Natural Sciences and Engineering Research Council of 
Canada; the National Science Council of the Republic of China; the 
Swiss National Science Foundation; the A.P. Sloan Foundation; the 
Bundesministerium f\"ur Bildung und Forschung, Germany; the Korean 
World Class University Program, the National Research Foundation of 
Korea; the Science and Technology Facilities Council and the Royal 
Society, United Kingdom; the Russian Foundation for Basic Research; 
the Ministerio de Ciencia e Innovaci\'{o}n, and Programa 
Consolider-Ingenio 2010, Spain; the Slovak R\&D Agency; the Academy of 
Finland; the Australian Research Council (ARC); and the EU community 
Marie Curie Fellowship Contract No. 302103. 
\end{acknowledgments}


\begin{thebibliography}{61}
\bibitem[]{CDFcoord}
CDF uses a cylindrical coordinate system with $+z$ in the proton beam direction. $\theta$ and
$\phi$ are the polar and azimuthal angles, respectively, and pseudorapidity
is $\eta = - \ln \tan(\theta/2)$.
\bibitem[]{EtDef}  
The transverse energy \et\ is defined as $E \sin\theta$.
\bibitem[]{IFSRUnc}
  We constrain the rate of initial-state radiation using Drell-Yan events in the data.
\end{thebibliography}
\end{document}